\documentclass[a4paper]{spie}
\usepackage[latin9]{inputenc}
\usepackage[english]{babel}
\usepackage{numprint}
\usepackage{graphicx}		
\usepackage[squaren,cdot]{SIunits}	

\title{Simulations of coronagraphy with a dynamic hologram \\ for the direct detection of exo-planets}
\author{
  Davide Ricci$^{1,}$\footnote{\ \ AEOS team, Scholar FRIA}, 
  Herv\'e Le Coroller$^2$, 
  Antoine Labeyrie$^3$,
  Pierre Piron$^1$\footnote{\ \ Laboratory HOLOLAB, Scholar FRIA} \\
  $^1$ D\'epartement d'Astrophysique, B\^at B5, Universit\'e de Li\`ege, B-4000 Li\`ege, (Belgium) \\
  $^2$ Observatoire de Haute Provence, F-04870 Saint Michel l'Observatoire (France) \\
  $^3$ Coll\`ege de France, 11, place Marcelin Berthelot F-75231 Paris Cedex 05 (France)
}

\begin{document}
\maketitle

\begin{abstract}
In a previous paper\cite{ricci}, we discussed an original solution to improve
the performances of coronagraphs by adding, in the optical scheme, an adaptive
hologram removing most of the residual speckle starlight.

\noindent In our simulations, the detection limit in the flux ratio between a
host star and a very near planet ($5 \lambda/D$) improves over a factor
\numprint{1000} (resp. \numprint{10000}) when equipped with a hologram for cases
of wavefront bumpiness imperfections of $\lambda/20$ (resp. $\lambda/100$).

\noindent We derive, in this paper, the transmission accuracy required on the
hologram pixels to achieve such goals. We show that preliminary tests could be
performed on the basis of existing technologies. 

\end{abstract}

\section{Introduction}

In the framework of the astronomical instrumentation for space-based telescopes,
coronagraphy plays a predominant role in the challenge for the direct detection
of extra-solar planets\cite{serabyn}. This technique, initially developed for
studying the solar corona masking the solar disk\cite{lyot}, evolved during the
last years replacing the initial opaque mask in the focal plane by various kinds
of phase masks\cite{riaud,roddier,rouand}, and/or apodizing the
pupil\cite{guyon}.

\noindent Anyway, even a perfect coronagraph is limited by the residual speckles
due to the bumpiness of the mirror and/or the imperfect AO.  The relative flux
between an exo-Earth and its parent star is typically $10^{-10}$.

\noindent Here, we present numerical simulations about the performances of this
coronagraph design, studied to overcome the previous
limitations\cite{labeyrie}. The star flux is recuperated instead of being
masked, and it is added, phase-shifted, to the coherent speckle halo, with the
help of a dynamic hologram placed on the relayed pupil plane. This operation
allows to remove most of the residual star light and increases the detectability
of a faint planet.

\noindent Our simulations refine the tests on this design proposed by our
group\cite{labeyrie,ricci}, by introducing a transmission noise on the
hologram. This will help us to understand the real performances of the hologram
in the framework of a practical implementation. Moreover, an estimation of the
error is given for all the simulations.

\noindent In Sect.~\ref{trad} we describe another optical design different from
the version proposed in our previous work\cite{ricci} in order to provide a
didactical explanation for implementing such a hologram in a lyot coronagraph.
In Sect.~\ref{numsim} we show the results of the numerical simulations, with
varying mirror bumpiness, and transmission noise on the hologram. We conclude in
Sect.~\ref{fine}.

\section{Traditional Lyot coronagraph and implementation with an adaptive hologram}
\label{trad}

\begin{figure}[t]
  \centering
  \includegraphics[width=0.70\textwidth]{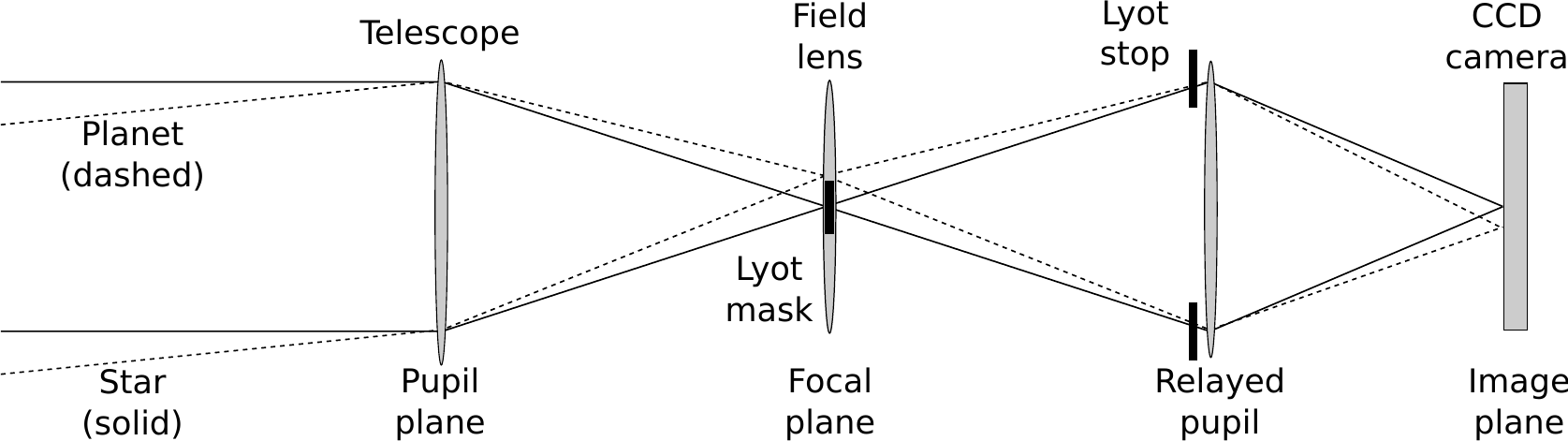}
  \caption{Classical Lyot coronagraph.
  }
  \label{lyot1}
\end{figure}

In a traditional coronagraph device (cfr. Fig.~\ref{lyot1}), the light of a star
and its companion is focused by the telescope in the focal plane, where an
opaque mask is placed. This element, called ``Lyot Mask'', has a typical size
designed to mask the light of the star without masking the light of the near
planet. A lens in the focal plane provides the re-imaging of the field in a
relayed pupil plane. Here, the residual light of the star is diffracted into a
bright ring. An annular opaque mask, called ``Lyot stop'', masks this light
improving the nulling of the star. Finally, a convergent lens re-images the
field  in its focal plane, where the detector (for example a CCD camera) is
placed.

\begin{figure}[t]
  \centering
  \includegraphics[width=0.70\textwidth]{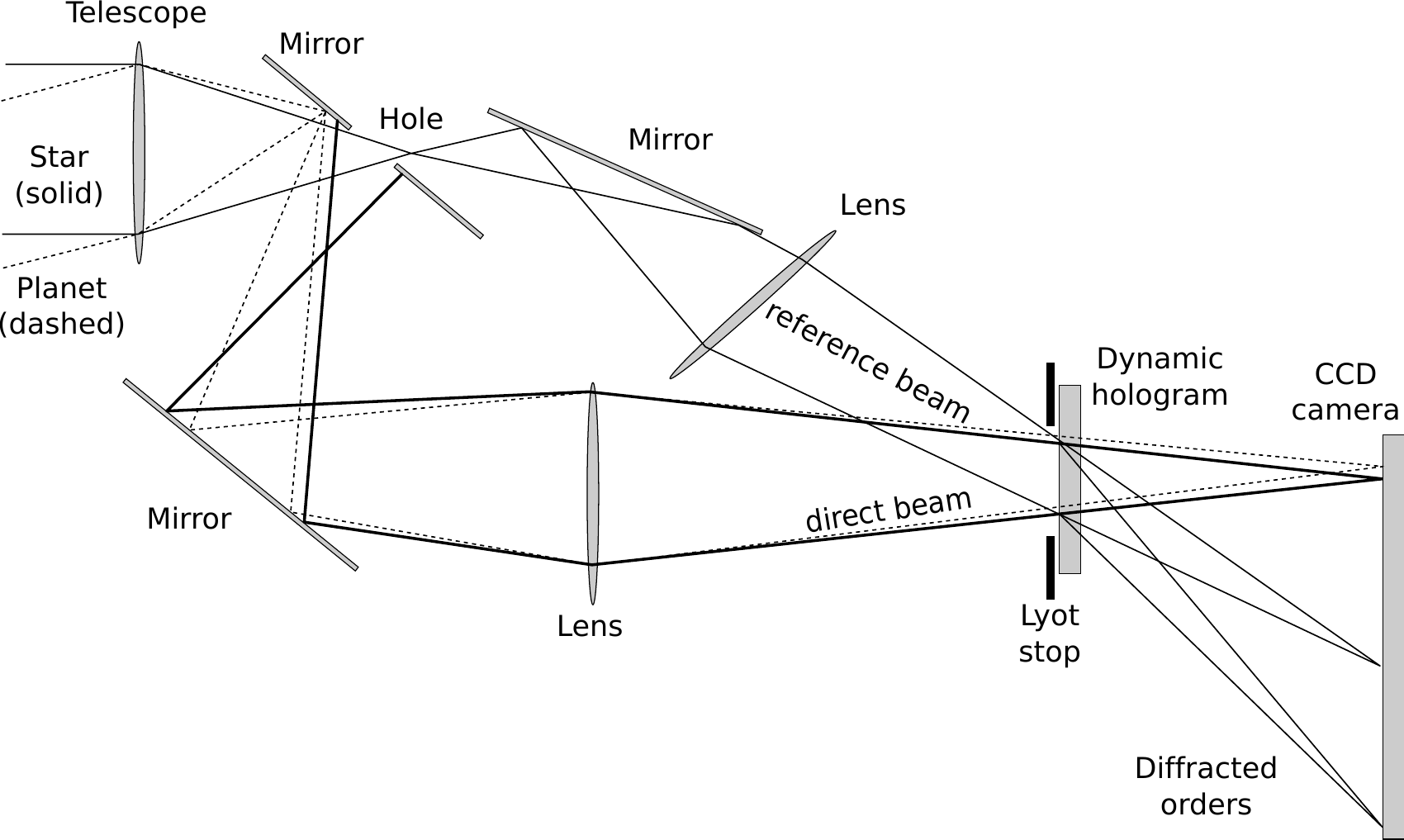}
  \caption{Implementation of an hologram in the classical Lyot coronagraph.
  }
  \label{lyot4}
\end{figure}

\noindent In our implementation (cfr. Fig.~\ref{lyot4}), the opaque mask is
replaced by a flat mirror with a central hole, so that the central part of the
beam, i.e. the light of the star, is not absorbed but it is recuperated. It is
referred to the ``reference beam''. The residual light of the star's Airy peak,
which does not pass through the hole, is referred to the ``direct beam'', and
proceeds its path together with the light of the planet. The reference beam is
deflected in such a way that it intersects the direct beam in the pupil plane,
where the Lyot stop is placed.

\noindent A dynamic hologram is located just behind the Lyot stop, and its
functioning is basically summarized as follows:

\begin{itemize}
\item 
  The reference beam and the direct beam add in a coherent way, creating fringes
  in the speckles of the hologram. This process is called ``recording the
  hologram''. The fringes of the recorded hologram act like a grating,
  diffracting several orders which become focused in the focal plane.
\item 
  by illuminating the recorded hologram with the reference beam only, a
  reconstructed image of the star's speckles appears on the detector, together
  with other diffracted orders.  Using a reference beam phase-shifted by $\pi$
  or a negative hologram, we obtain the same image phase-shifted by $\pi$.
\item 
  Our technique consists in creating an interference between the direct beam and
  the $\pi$ phase-shifted of the order +1 of the reference beam.  Thanks to this
  phase shifting, the order 0 of the direct beam adds destructively with the
  reconstructed image of the star's speckle, thus nulling the residual speckles
  of the star.
\item
  The planet's light, being incoherent with respect to the reference beam is not
  reconstructed by the hologram, and not erased in the focal plane.
\end{itemize}

\noindent In a more performing design, the hole in the flat mirror can be
replaced by a micro-prism.  This approach is presented in our recent
paper\cite{ricci}, while here we have presented a simpler concept.

\section{Numerical simulations}
\label{numsim}

First, we have re-computed a large number of simulations presented in our paper
in order to display graphics with error bars: we simulated a $6.5\meter$ space
telescope (analogue to the James Webb Space Telescope, hereafter JWST), equipped
with our coronagraphic system, and we imagined the telescope observing an
analogue of the Sun-Earth system at $11\rm pc$ from us. From that position, the
angular separation between the star and the Earth-like planet would be such that
the planet would lie on the fifth ring of the Airy pattern created by his parent
star. 

\noindent In our setup we maintained the optimizations already
determined\cite{ricci}, and we tested the performances under different
conditions. Each simulation was calculated under the assumption of mirror bumpiness
of $\lambda/20$ and $\lambda/100$.
\begin{figure}[p]
  \centering
  \includegraphics[width=0.48\textwidth]{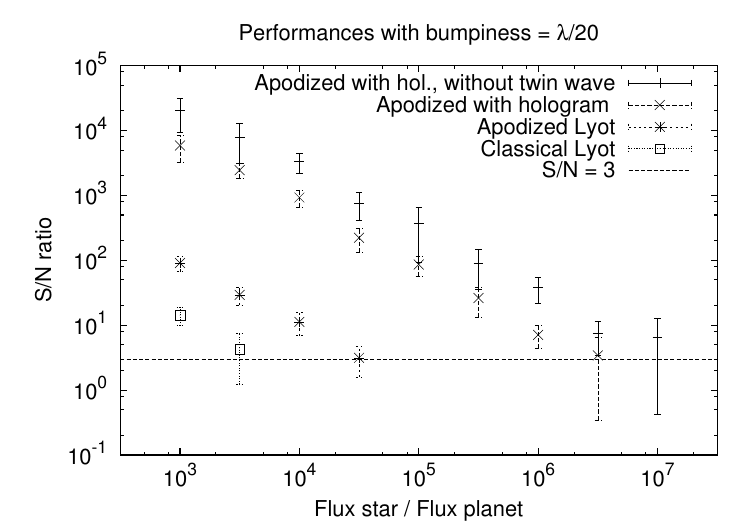}
  \includegraphics[width=0.48\textwidth]{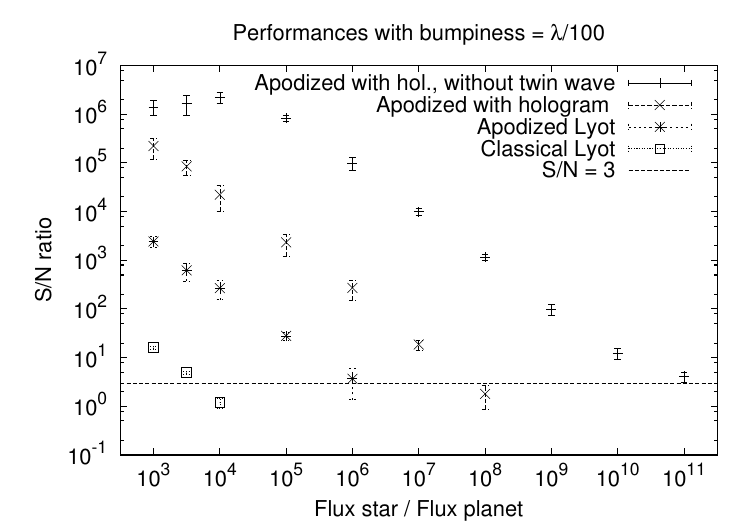}
  \caption{Simulations of several kinds of coronagraphs (see the legend). The
    figures show the signal-to-noise ratio (S/N) as a function of the ratio between
    the flux of the host star and the flux of the planet for the cases of mirror
    bumpiness of $\lambda/20$ (left) and $\lambda/100$ (right).  }
  \label{coro}
\end{figure}
\begin{figure}[p]
  \centering
  \includegraphics[width=0.48\textwidth]{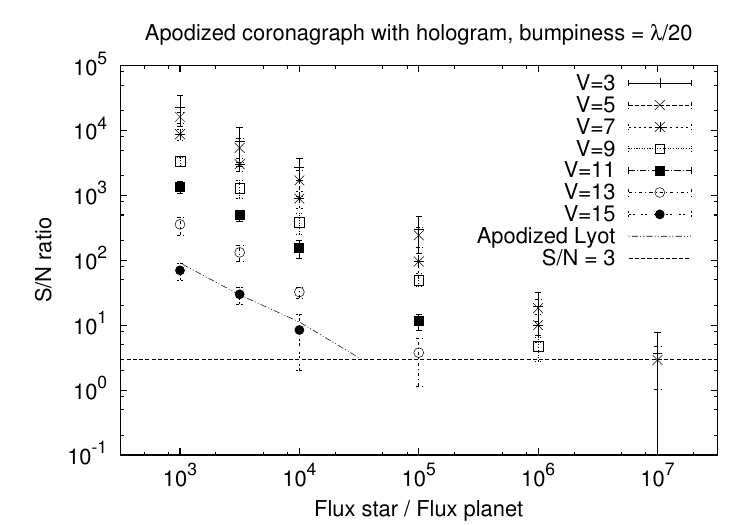}
  \includegraphics[width=0.48\textwidth]{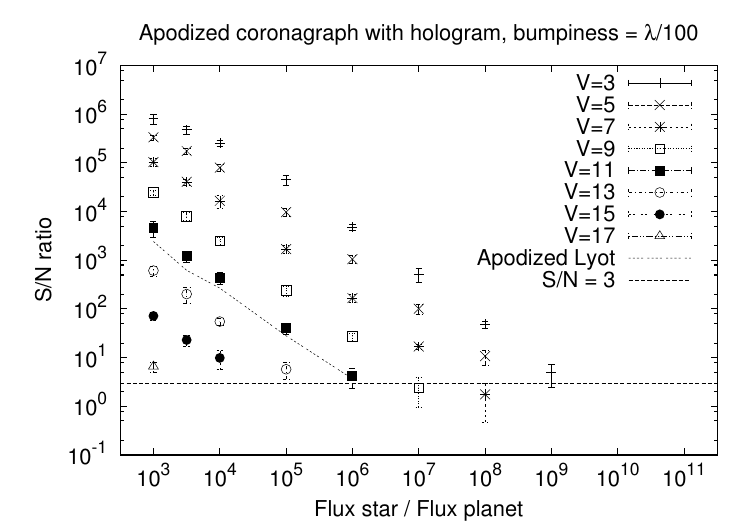}
  \caption{Simulation of an apodized Lyot coronagraph equipped with a dynamic
    hologram, for different $V$ magnitudes (see legend), in the case of noise
    both on the hologram and on the CCD device.  The figures show the S/N as a
    function of the ratio between the flux of the host star and the flux of the
    planet for the cases of mirror bumpiness of $\lambda/20$ (left) and
    $\lambda/100$ (right), with a JWST-like space telescope and with an exposure
    time of $60\second$. }
  \label{olo}
\end{figure}
\begin{figure}[p]
  \centering
  \includegraphics[width=0.48\textwidth]{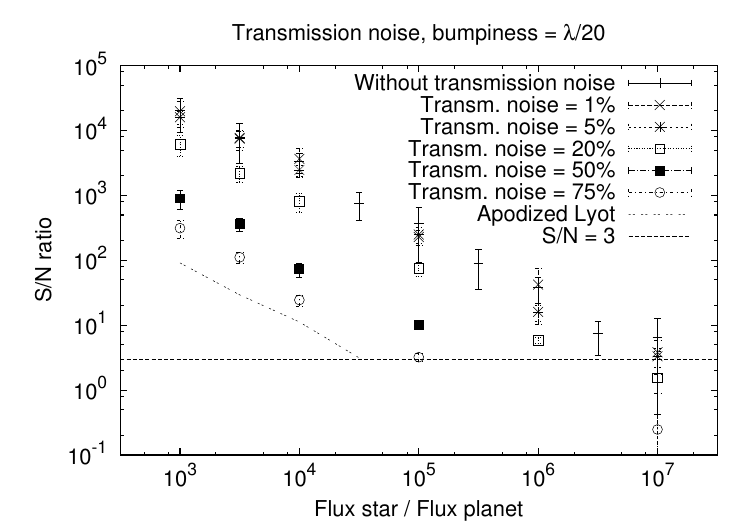}
  \includegraphics[width=0.48\textwidth]{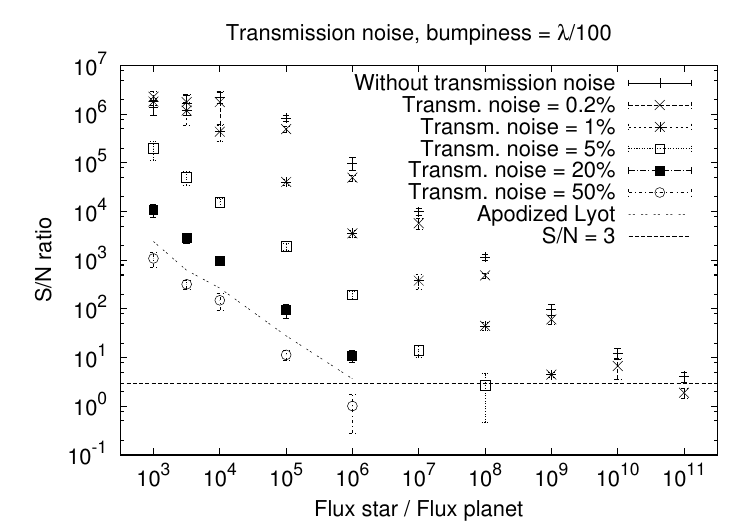}
  \caption{Simulations of an apodized Lyot coronagraph equipped with a dynamic
    hologram, without photon noise, for different amplitudes of transmission
    noise on the hologram. The figures show the S/N as a function of the ratio
    between the flux of the host star and the flux of the planet for the cases
    of mirror bumpiness of $\lambda/20$ (left) and $\lambda/100$ (right).  }
  \label{trans}
\end{figure}
\begin{itemize}
\item
  Simulations without photon noise. Several kinds of coronagraphs were
  simulated: a classical Lyot coronagraph; a Lyot coronagraph with apodized
  pupil; an apodized coronagraph provided with hologram; and finally an apodized
  coronagraph provided with hologram and with the analytical subtraction of the
  twin wave (see our recent work\cite{ricci} for more explanation about the twin
  wave). Under these conditions, the magnitude of the star is indifferent, as
  the absence of photon noise does not degrade the signal-to-noise ratio. The
  results are shown in Fig.~\ref{coro}. \\ Under such condition we are able to
  increase the performances by a factor $10^3$ (at $\lambda / 20$) and $10^5$
  (at $\lambda / 100$) with respect to the apodized coronagraph.
\item
  Simulations adding photon noise both on the final image (CCD) and on the
  hologram. We concentrated on the differences between our most performing
  coronagraph (apodized, provided with hologram and without twin wave) and its
  analogue without hologram (the apodized coronagraph). In this situation it is
  possible to test the limits of validity of our instrument as a function of the
  magnitude of the star.  The results are shown in Fig.~\ref{olo}. \\ When the
  star is bright enough, we reach the maximum nulling power achievable with an
  hologram: this happens at magnitude $V=7$ (at $\lambda / 20$) and $V=3$ (at
  $\lambda / 100$), and improves the detection limit by a factor $10^3$ and
  $10^4$ respectively.  When the star is too faint, the performances are not
  better than those of the apodized coronagraph: this happens at magnitude
  $V=15$ (at $\lambda / 20$) and $V=11$ (at $\lambda / 100$).
\item
  Simulations adding transmission noise on the hologram.
  We simulated several levels of random transmission noises to
  create these transmission errors on the hologram pixels.  The results are
  shown in Fig.~\ref{trans}. \\ The effect of this kind of noise is numerically
  approximately equivalent to the photon noise treated above (even if it is not
  the same phenomenon), and once again we show the limits of validity of our
  instrument, by varying the amount of the transmission noise: when this noise
  is less than $5\%$ of the value of the hologram pixels (at $\lambda / 20$) and
  $0.2\%$ (at $\lambda / 100$), the performances are not degraded. If the
  transmission noise is more than $20\%$ of the value of the hologram pixels (at
  $\lambda / 20$) and $75\%$ (at $\lambda / 100$), the performances match those
  of the apodized coronagraph.
\end{itemize}
\noindent Several sets of measurements were performed, in order to obtain
different realizations of  noise and calculate for the error (rms of the mean
of the values at the $1\sigma$ level).

\noindent We performed a preliminary laboratory test using a static 2-layer
hologram (a photo-polymerizable layer and a liquid crystal polymer layer). 
The results are encouraging but further tests are required.

\section{Conclusions}
\label{fine}

Our simulations show that it is possible to override the current limits of
coronagraphic systems by adding, in the optical scheme, a dynamic hologram able
to remove the residual speckle starlight.

\noindent The tests presented in this paper indicate that, in the framework of a
practical implementation, it is sufficient to control the ``actuators'' of the
hologram with a precision better than $0.2$--$5\%$ in order not to be limited by
the transmission noise. This allows to improve the performances of the coronagraph
by a factor \numprint{1000}--\numprint{10000}, and make this technique suitable 
for being developed and successfully applied to the next generation of space 
telescopes.

\noindent In the near furure, we will experiment such a coronagraphic hologram
on a laboratory test bench.

\bibliographystyle{spiebib}
\bibliography{biblio}

\end{document}